\documentclass{article}
\usepackage{spconf}

\usepackage{amssymb, amsmath}
\usepackage{amsfonts}
\usepackage{bbm}
\usepackage{mathrsfs}
\usepackage{xspace}
\usepackage{bm}

\usepackage{cite}
\usepackage{epsfig}
\usepackage{cases}
\usepackage{psfrag}
\usepackage{enumerate}

\usepackage{makeidx}
\usepackage{supertabular}

\newtheorem{theorem}{Theorem}

\newtheorem{lemma}{Lemma}

\newenvironment{textbmatrix}{   \setlength{\arraycolsep}{2.5pt}%
                                                                \big[\begin{matrix}}{\end{matrix}\big]%
                                                                \raisebox{0.08ex}{\vphantom{M}}}

\def\be{\begin{equation}}
\def\ee{\end{equation}}
\def\een{\nonumber \end{equation}}
\def\mat{\begin{bmatrix}}
\def\emat{\end{bmatrix}}
\def\btm{\begin{textbmatrix}}
\def\etm{\end{textbmatrix}}

\def\ba#1\ea{\begin{align}#1\end{align}}
\def\bs#1\es{\begin{split}#1\end{split}} 
\def\bg#1\eg{\begin{gather}#1\end{gather}} 
\def\bi#1\ei{\begin{itemize}#1\end{itemize}}

\newcommand{\safemath}[2]{\newcommand{#1}{\ensuremath{#2}\xspace}}

\safemath{\interior}{\mathrm{Int}}                       

\safemath{\dfn}{:=}                                                     
\safemath{\dirac}{\delta}                                       

\safemath{\SNR}{\text{\sc snr}}                                 
\safemath{\No}{N_0}                                                     
\safemath{\Es}{E_s}                                                     
\safemath{\Eb}{E_b}                                                     
\safemath{\EbNo}{\frac{\Eb}{\No}}
\safemath{\EsNo}{\frac{\Es}{\No}}

\DeclareMathOperator{\CHop}{\ensuremath{\mathbb{H}}} 
\safemath{\tvir}{h_{\CHop}}                                     
\safemath{\tvtf}{L_{\CHop}}                                     
\safemath{\spf}{S_{\CHop}}                                              
\safemath{\bff}{H_{\CHop}}                                      

\safemath{\ircf}{R_{h}}                                         
\safemath{\scf}{R_{S}}                                          
\safemath{\tfcf}{R_{L}}                                         
\safemath{\bfcf}{R_{H}}                                         

\safemath{\mi}{I}                                                       
\safemath{\capacity}{C}                                         

\newcommand{\pdf[2]}{f_{{#1}}\left({#2}\right)}                    

\safemath{\uniform}{\mathcal{U}}                        
\safemath{\normal}{\mathcal{N}}                         
\safemath{\circnorm}{\mathcal{CN}}                      
\safemath{\mchain}{\leftrightarrow}                     

\safemath{\dB}{\,\mathrm{dB}}
\safemath{\dBm}{\,\mathrm{dBm}}
\safemath{\Hz}{\,\mathrm{Hz}}
\safemath{\kHz}{\,\mathrm{kHz}}
\safemath{\MHz}{\,\mathrm{MHz}}
\safemath{\GHz}{\,\mathrm{GHz}}
\safemath{\s}{\,\mathrm{s}}
\safemath{\ms}{\,\mathrm{ms}}
\safemath{\mus}{\,\mathrm{\mu s}}
\safemath{\ns}{\,\mathrm{ns}}
\safemath{\meter}{\,\mathrm{m}}
\safemath{\km}{\,\mathrm{km}}
\safemath{\mm}{\,\mathrm{mm}}
\safemath{\cm}{\,\mathrm{cm}}
\safemath{\m}{\,\mathrm{m}}
\safemath{\W}{\,\mathrm{W}}
\safemath{\J}{\,\mathrm{J}}
\safemath{\K}{\,\mathrm{K}}
\safemath{\bit}{\,\mathrm{bit}}
\safemath{\nW}{\,\mathrm{nW}}
\safemath{\muW}{\,\mathrm{$\mu$W}}
\safemath{\Watt}{\,\mathrm{W}}

\safemath{\define}{\triangleq}                  

\safemath{\equivalent}{\sim}
\safemath{\distas}{\sim}                                        

\safemath{\reals}{\mathbb{R}}
\safemath{\positivereals}{\mathbb{R}^{+}}
\safemath{\integers}{\mathbb{Z}}
\safemath{\posint}{\mathbb{Z}_{+}}
\safemath{\naturals}{\mathbb{N}}
\safemath{\complexset}{\mathbb{C}}

\safemath{\setA}{\mathcal{A}}
\safemath{\setB}{\mathcal{B}}
\safemath{\setC}{\mathcal{C}}
\safemath{\setD}{\mathcal{D}}
\safemath{\setE}{\mathcal{E}}
\safemath{\setF}{\mathcal{F}}
\safemath{\setG}{\mathcal{G}}
\safemath{\setH}{\mathcal{H}}
\safemath{\setI}{\mathcal{I}}
\safemath{\setJ}{\mathcal{J}}
\safemath{\setK}{\mathcal{K}}
\safemath{\setL}{\mathcal{L}}
\safemath{\setM}{\mathcal{M}}
\safemath{\setN}{\mathcal{N}}
\safemath{\setO}{\mathcal{O}}
\safemath{\setP}{\mathcal{P}}
\safemath{\setQ}{\mathcal{Q}}
\safemath{\setR}{\mathcal{R}}
\safemath{\setS}{\mathcal{S}}
\safemath{\setT}{\mathcal{T}}
\safemath{\setU}{\mathcal{U}}
\safemath{\setV}{\mathcal{V}}
\safemath{\setW}{\mathcal{W}}
\safemath{\setX}{\mathcal{X}}
\safemath{\setY}{\mathcal{Y}}
\safemath{\setZ}{\mathcal{Z}}
\safemath{\emptySet}{\varnothing}

\safemath{\bma}{\mathbf{a}}
\safemath{\bmb}{\mathbf{b}}
\safemath{\bmc}{\mathbf{c}}
\safemath{\bmd}{\mathbf{d}}
\safemath{\bme}{\mathbf{e}}
\safemath{\bmf}{\mathbf{f}}
\safemath{\bmg}{\mathbf{g}}
\safemath{\bmh}{\mathbf{h}}
\safemath{\bmi}{\mathbf{i}}
\safemath{\bmj}{\mathbf{j}}
\safemath{\bmk}{\mathbf{k}}
\safemath{\bml}{\mathbf{l}}
\safemath{\bmm}{\mathbf{m}}
\safemath{\bmn}{\mathbf{n}}
\safemath{\bmo}{\mathbf{o}}
\safemath{\bmp}{\mathbf{p}}
\safemath{\bmq}{\mathbf{q}}
\safemath{\bmr}{\mathbf{r}}
\safemath{\bms}{\mathbf{s}}
\safemath{\bmt}{\mathbf{t}}
\safemath{\bmu}{\mathbf{u}}
\safemath{\bmv}{\mathbf{v}}
\safemath{\bmw}{\mathbf{w}}
\safemath{\bmx}{\mathbf{x}}
\safemath{\bmy}{\mathbf{y}}
\safemath{\bmz}{\mathbf{z}}

\bmdefine{\biad}{a}
\bmdefine{\bibd}{b}
\bmdefine{\bicd}{c}
\bmdefine{\bidd}{d}
\bmdefine{\bied}{e}
\bmdefine{\bifd}{f}
\bmdefine{\bigd}{g}
\bmdefine{\bihd}{h}
\bmdefine{\biid}{i}
\bmdefine{\bijd}{j}
\bmdefine{\bikd}{k}
\bmdefine{\bild}{l}
\bmdefine{\bimd}{m}
\bmdefine{\bind}{n}
\bmdefine{\biod}{o}
\bmdefine{\bipd}{p}
\bmdefine{\biqd}{q}
\bmdefine{\bird}{r}
\bmdefine{\bisd}{s}
\bmdefine{\bitd}{t}
\bmdefine{\biud}{u}
\bmdefine{\bivd}{v}
\bmdefine{\biwd}{w}
\bmdefine{\bixd}{x}
\bmdefine{\biyd}{y}
\bmdefine{\bizd}{z}

\bmdefine{\bixid}{\xi}
\bmdefine{\bilambdad}{\lambda}
\bmdefine{\bimud}{\mu}
\bmdefine{\bithetad}{\theta}
\bmdefine{\biphid}{\phi}

\safemath{\bmia}{\biad}
\safemath{\bmib}{\bibd}
\safemath{\bmic}{\bicd}
\safemath{\bmid}{\bidd}
\safemath{\bmie}{\bied}
\safemath{\bmif}{\bifd}
\safemath{\bmig}{\bigd}
\safemath{\bmih}{\bihd}
\safemath{\bmii}{\biid}
\safemath{\bmij}{\bijd}
\safemath{\bmik}{\bikd}
\safemath{\bmil}{\bild}
\safemath{\bmim}{\bimd}
\safemath{\bmin}{\bind}
\safemath{\bmio}{\biod}
\safemath{\bmip}{\bipd}
\safemath{\bmiq}{\biqd}
\safemath{\bmir}{\bird}
\safemath{\bmis}{\bisd}
\safemath{\bmit}{\bitd}
\safemath{\bmiu}{\biud}
\safemath{\bmiv}{\bivd}
\safemath{\bmiw}{\biwd}
\safemath{\bmix}{\bixd}
\safemath{\bmiy}{\biyd}
\safemath{\bmiz}{\bizd}

\safemath{\bmxi}{\bixid}
\safemath{\bmlambda}{\bilambdad}
\safemath{\bmmu}{\bimud}
\safemath{\bmtheta}{\bithetad}
\safemath{\bmphi}{\biphid}

\safemath{\bA}{\mathbf{A}}
\safemath{\bB}{\mathbf{B}}
\safemath{\bC}{\mathbf{C}}
\safemath{\bD}{\mathbf{D}}
\safemath{\bE}{\mathbf{E}}
\safemath{\bF}{\mathbf{F}}
\safemath{\bG}{\mathbf{G}}
\safemath{\bH}{\mathbf{H}}
\safemath{\bI}{\mathbf{I}}
\safemath{\bJ}{\mathbf{J}}
\safemath{\bK}{\mathbf{K}}
\safemath{\bL}{\mathbf{L}}
\safemath{\bM}{\mathbf{M}}
\safemath{\bN}{\mathbf{N}}
\safemath{\bO}{\mathbf{O}}
\safemath{\bP}{\mathbf{P}}
\safemath{\bQ}{\mathbf{Q}}
\safemath{\bR}{\mathbf{R}}
\safemath{\bS}{\mathbf{S}}
\safemath{\bT}{\mathbf{T}}
\safemath{\bU}{\mathbf{U}}
\safemath{\bV}{\mathbf{V}}
\safemath{\bW}{\mathbf{W}}
\safemath{\bX}{\mathbf{X}}
\safemath{\bY}{\mathbf{Y}}
\safemath{\bZ}{\mathbf{Z}}

\safemath{\bZero}{\mathbf{0}}

\bmdefine{\biAd}{A}
\bmdefine{\biBd}{B}
\bmdefine{\biCd}{C}
\bmdefine{\biDd}{D}
\bmdefine{\biEd}{E}
\bmdefine{\biFd}{F}
\bmdefine{\biGd}{G}
\bmdefine{\biHd}{H}
\bmdefine{\biId}{I}
\bmdefine{\biJd}{J}
\bmdefine{\biKd}{K}
\bmdefine{\biLd}{L}
\bmdefine{\biMd}{M}
\bmdefine{\biOd}{N}
\bmdefine{\biPd}{O}
\bmdefine{\biQd}{P}
\bmdefine{\biRd}{R}
\bmdefine{\biSd}{S}
\bmdefine{\biTd}{T}
\bmdefine{\biUd}{U}
\bmdefine{\biVd}{V}
\bmdefine{\biWd}{W}
\bmdefine{\biXd}{X}
\bmdefine{\biYd}{Y}
\bmdefine{\biZd}{Z}

\bmdefine{\biDelta}{\Delta}
\bmdefine{\biLambda}{\Lambda}
\bmdefine{\biPhi}{\Phi}
\bmdefine{\biSigma}{\Sigma}
\bmdefine{\biOmega}{\Omega}
\bmdefine{\biTheta}{\Theta}

\safemath{\bimA}{\biAd}
\safemath{\bimB}{\biBd}
\safemath{\bimC}{\biCd}
\safemath{\bimD}{\biDd}
\safemath{\bimE}{\biEd}
\safemath{\bimF}{\biFd}
\safemath{\bimG}{\biGd}
\safemath{\bimH}{\biHd}
\safemath{\bimI}{\biId}
\safemath{\bimJ}{\biJd}
\safemath{\bimK}{\biKd}
\safemath{\bimL}{\biLd}
\safemath{\bimM}{\biMd}
\safemath{\bimN}{\biNd}
\safemath{\bimO}{\biOd}
\safemath{\bimP}{\biPd}
\safemath{\bimQ}{\biQd}
\safemath{\bimR}{\biRd}
\safemath{\bimS}{\biSd}
\safemath{\bimT}{\biTd}
\safemath{\bimU}{\biUd}
\safemath{\bimV}{\biVd}
\safemath{\bimW}{\biWd}
\safemath{\bimX}{\biXd}
\safemath{\bimY}{\biYd}
\safemath{\bimZ}{\biZd}

\safemath{\bDelta}{\bielta}
\safemath{\bLambda}{\biLambda}
\safemath{\bPhi}{\biPhi}
\safemath{\bSigma}{\biSigma}
\safemath{\bOmega}{\biOmega}
\safemath{\bTheta}{\biTheta}

\safemath{\veca}{\bma}
\safemath{\vecb}{\bmb}
\safemath{\vecc}{\bmc}
\safemath{\vecd}{\bmd}
\safemath{\vece}{\bme}
\safemath{\vecf}{\bmf}
\safemath{\vecg}{\bmg}
\safemath{\vech}{\bmh}
\safemath{\veci}{\bmi}
\safemath{\vecj}{\bmj}
\safemath{\veck}{\bmk}
\safemath{\vecl}{\bml}
\safemath{\vecm}{\bmm}
\safemath{\vecn}{\bmn}
\safemath{\veco}{\bmo}
\safemath{\vecp}{\bmp}
\safemath{\vecq}{\bmq}
\safemath{\vecr}{\bmr}
\safemath{\vecs}{\bms}
\safemath{\vect}{\bmt}
\safemath{\vecu}{\bmu}
\safemath{\vecv}{\bmv}
\safemath{\vecw}{\bmw}
\safemath{\vecx}{\bmx}
\safemath{\vecy}{\bmy}
\safemath{\vecz}{\bmz}
\safemath{\vecZero}{\bZero}

\safemath{\vecxi}{\bmxi}
\safemath{\veclambda}{\bmlambda}
\safemath{\vecmu}{\bmmu}
\safemath{\vectheta}{\bmtheta}
\safemath{\vecphi}{\bmphi}

\safemath{\matA}{\bA}
\safemath{\matB}{\bB}
\safemath{\matC}{\bC}
\safemath{\matD}{\bD}
\safemath{\matE}{\bE}
\safemath{\matF}{\bF}
\safemath{\matG}{\bG}
\safemath{\matH}{\bH}
\safemath{\matI}{\bI}
\safemath{\matJ}{\bJ}
\safemath{\matK}{\bK}
\safemath{\matL}{\bL}
\safemath{\matM}{\bM}
\safemath{\matN}{\bN}
\safemath{\matO}{\bO}
\safemath{\matP}{\bP}
\safemath{\matQ}{\bQ}
\safemath{\matR}{\bR}
\safemath{\matS}{\bS}
\safemath{\matT}{\bT}
\safemath{\matU}{\bU}
\safemath{\matV}{\bV}
\safemath{\matW}{\bW}
\safemath{\matX}{\bX}
\safemath{\matY}{\bY}
\safemath{\matZ}{\bZ}
\safemath{\matZero}{\bZero}

\safemath{\matDelta}{\bDelta}
\safemath{\matLambda}{\bLambda}
\safemath{\matPhi}{\bPhi}
\safemath{\matSigma}{\bSigma}
\safemath{\matOmega}{\bOmega}
\safemath{\matTheta}{\bTheta}

\safemath{\matIdentity}{\matI}

\usepackage{algorithm,algorithmic}
\usepackage[usenames, dvipsnames]{color}

\safemath{\crb}{\mathsf{CRB}}
\safemath{\mse}{\mathsf{MSE}}
\safemath{\var}{\mathsf{var}}
\safemath{\nvar}{\mathsf{nvar}}

\safemath{\dif}{\,\mathrm{d}}

\newfont{\bb}{msbm10 scaled 1100}

\usepackage[font=small,skip= -10pt]{caption}

\begin{document}

\title{Convex separable problems with linear and box constraints}

\name{\vspace{-.4cm}Antonio A. D'Amico$^{\star}$ \qquad Luca Sanguinetti$^{\star \dagger}$ \qquad Daniel P. Palomar$^{\ddagger}$ \thanks{L. Sanguinetti has received funding from the People Programme (Marie Curie Actions) of the FP7 PIEF-GA-2012-330731 Dense4Green and from the FP7 Network of Excellence in Wireless COMmunications NEWCOM\# (Grant agreement no. 318306). Daniel P. Palomar has been supported by the Hong Kong RPC11EG39 research grant.}}
\address{$^{\star}$\small{Dipartimento di Ingegneria dell'Informazione, University of Pisa, Pisa, Italy}\\
$^{\dagger}$\small{Alcatel-Lucent Chair, Ecole sup{\'e}rieure d'{\'e}lectricit{\'e} (Sup{\'e}lec), Gif-sur-Yvette, France}\\
$^\ddagger$\small{Department of Electronic and Computer Engineering, University of Science and Technology,
Hong Kong}}

  \maketitle
\vspace{-.5cm}
\begin{abstract}
In this work, we focus on separable convex optimization problems with linear and box constraints and compute the solution in closed-form as a function of some Lagrange multipliers that can be easily computed in a finite number of iterations. This allows us to bridge the gap between a wide family of power allocation problems of practical interest in signal processing and communications and their efficient implementation in practice.
\end{abstract}

\vspace{-.4cm}
\section{Introduction}
\vspace{-.2cm}
Consider the following problem:\vspace{-.2cm}
\begin{align}\label{1}
     (\mathcal P): \quad  \underset{\{x_n\}}{\min} \quad &
\sum\limits_{n=1}^N f_n(x_n)\\\nonumber
\text{subject to} \quad & \sum \limits_{n=1}^{j} x_n\le \rho_j \quad  j=1,\ldots,N\\\nonumber
 & l_n \le x_n \le u_n \quad  n=1,\ldots,N
\end{align}
where $\{x_n\}$ are the optimization variables, the coefficients $\{\rho_j\}$ are real-valued parameters whereas the constraints $ l_n \le x_n \le u_n$ are called variable bounds or box constraints with $-\infty \le l_n < u_n \le +\infty$. The functions $f_{n}$ are real-valued, continuous and strictly convex in $[l_n,u_{n}]$, and continuously differentiable in $(l_n,u_{n})$. If $f_{n}$ is not defined in $l_n$ and/or $u_n$, then it is extended by continuity assuming $f_n(l_n)=\mathop {\lim }\nolimits_{x_n \rightarrow l_n^+} f_n(x_n)$ and $f_n(u_n)=\mathop {\lim }\nolimits_{x_n \rightarrow u_n^-} f_n(x_n)$.

Constrained optimization problems in the form given by $(\mathcal P)$ in \eqref{1} arise in connection with a wide range of applications and settings in signal processing and communications. For example, they arise in connection with the design of multiple-input multiple-output (MIMO) systems dealing with the minimization of the power consumption while meeting the quality-of-service (QoS) requirements over each data stream (see for example \cite{PalomarQoS2004} -- \cite{Jiang2006} for point-to-point communications and \cite{Fu2011} -- \cite{Sanguinetti2012} for two-hop amplify-and-forward relay networks). A good survey of some of these problems for point-to-point MIMO communications can be found in \cite{PalomarWF}. They also appear in the design of optimal training sequences for channel estimation in multi-hop transmissions using decode-and-forward protocols \cite{Gao2008} and in the optimal power allocation for the maximization of the instantaneous received signal-to-noise ratio in amplify-and-forward multi-hop transmissions under short-term power constraints \cite{Farhadi09}.
Other instances of \eqref{1} are shown to be the rate-constrained power minimization problem over a code division multiple-access channel with correlated noise \cite{Padakandla2009} and the power allocation problem in amplify-and-forward relaying scheme for multiuser cooperative networks under frequency-selective block-fading \cite{Pham2010}.

Clearly, the optimization problem in \eqref{1} can always be solved using standard convex solvers. Although possible, this in general does not provide any insights into its solution and does not exploit the particular structure of the problem itself. In this respect, all the aforementioned works go a step further and provide ad-hoc algorithms for their specific problems at hand in the attempt of giving some intuition on the solutions. The main contribution of this work is to develop a general framework that allows one to compute the solution (and its structure) for any problem in the form of \eqref{1}. This allows us to bridge the gap between a wide family of problems in signal processing and communications and their implementation in practice. In other words, whenever a problem can be put in the form of \eqref{1}, then its solution can be efficiently obtained by particularizing the proposed algorithm to the problem at hand.

The main related literature to this paper is represented by \cite{PalomarWF}, \cite{Padakandla2007} and \cite{Wang2012}. In \cite{PalomarWF}, the authors propose a general framework for solving optimization problems in which the solutions have a waterfilling structure. In \cite{Padakandla2007} and \cite{Wang2012}, the authors focus on solving problems of the form:\vspace{-.2cm}
\begin{align}\label{1.10}
 \underset{\{x_n\}}{\min} \quad &
\sum\limits_{n=1}^N f_n(x_n)\\\nonumber
\text{subject to} \quad & \sum \limits_{n=1}^{j} x_n\le \sum\limits_{n=1}^j\alpha_n \quad  j=1,\ldots,N\\\nonumber
 & 0 \le x_n \le u_n \quad  n=1,\ldots,N
\end{align}
with $\alpha_n \ge 0$ $\forall n$, which are known as separable convex optimization problems with linear ascending inequality constraints and box constraints. In particular, in \cite{Padakandla2007} the authors propose a dual method to numerically evaluate the solution of the above problem in no more than $N-1$ iterations under an ordering condition on the slopes of the functions at the origin. An alternative solution improving the worst case complexity of \cite{Padakandla2007} is illustrated in \cite{Wang2012}. Differently from \cite{Padakandla2007} and \cite{Wang2012}, we consider more general problems in which the inequality constraints are not necessarily in ascending order and the box constraint values $l_n$ and $u_n$ may possibly be equal to $-\infty$ and $+\infty$, respectively. All this makes \eqref{1} more general than problems of the form given in \eqref{1.10}. Also, some of the restrictions on functions $f_n$ that were present in \cite{Padakandla2007} and \cite{Wang2012} have been removed\footnote{It is also worth mentioning that at the time of submission we became aware of \cite{Akhil2014} in which the authors come up with an extended solution much similar to the proposed one using the theory of polymatroids.}.

\vspace{-.5cm}
\section{Preliminary results}\vspace{-.2cm}
We begin by denoting $\{x^{\star}_n;n=1,\ldots,N\}$ the solutions of \eqref{1} and observing that a necessary and sufficient condition for \eqref{1} to be feasible is easily given by 
\begin{align}\label{feasibility}
\sum \limits_{n=1}^{j} l_n\le \rho_j \quad  j=1,\ldots,N.
\end{align}
In addition, we observe that since $f_n$ is by definition continuous and strictly convex in $[l_n,u_{n}]$, and continuously differentiable in $(l_n,u_{n})$, then the three following cases may occur:

{\bf{a}}) The function $f_n$ is monotonically increasing in $[l_n,u_{n}]$ or, equivalently, $f_n'(x_n)>0$ for any $x_n \in (l_n,u_{n})$.

{\bf{b}}) There exists a point $z_n$ in $(l_n,u_n)$ such that $f_n^{'} (z_n) = 0 $ with $f_n'(x_n)<0$ and $f_n'(x_n)>0$ for any $x_n$ in $(l_n, z_n)$ and $(z_n, u_n)$, respectively.

{\bf{c}}) The function $f_n$ is monotonically decreasing in $[l_n,u_{n}]$ or, equivalently, $f_n'(x_n)<0$ for any $x_n \in (l_n,u_{n})$.

    \begin{lemma} \label{LemmaA}
    If $f_n$ is monotonically increasing in $[l_n,u_{n}]$ and $l_n \ne - \infty$, then $x_n^\star$ is given by $x_n^\star  = l_n$.
    \end{lemma}

The proof of the above lemma can be found in \cite{Sanguinetti2013} and can be used to find an equivalent form of \eqref{1}. To see how this comes about, denote by $\mathcal A \subseteqq \mathcal \{1,\ldots,N\}$ the set of indices $n$ in \eqref{1} for which case {\bf{a}}) holds true and assume (without loss of generality)
that $\mathcal A=\{1,2,\ldots,|\mathcal A|\}$.
Using the results of Lemma 1, it follows that $x_n^\star  = l_n$ for any $n \in \mathcal A$ while the computation of the remaining variables with indices $n \notin \mathcal A$ requires to solve the following reduced problem:\vspace{-.2cm}
    \begin{align}\label{caseA}
     \underset{\{x_n\}}{\min} \quad &
                                      \sum\limits_{n=|\mathcal A|+1}^{N} f_n(x_n)\\\nonumber
     \text{subject to} \quad & \sum \limits_{n=|\mathcal A|+1}^{j} x_n\le \rho_j^{\prime} \quad  j=|\mathcal A|+1,\ldots,N\\\nonumber
     & l_n \le x_n \le u_n \quad  n=|\mathcal A|+1,\ldots,N
    \end{align}
    with $\rho_j^{\prime} = \rho_j-\sum\nolimits_{n=1}^{|\mathcal A|}l_n$
for $ j = |\mathcal A|+1,\ldots,N$\footnote{Notice that in order for problem in \eqref{caseA} and thus for the original problem in \eqref{1} to be well-defined it must be $l_n \ne - \infty$ $\forall n \in \mathcal A$.}. The above optimization problem is exactly in the same form of \eqref{1} except for the fact that all its functions $f_n$ fall into cases {\bf{b}}) or {\bf{c}}). To proceed further, we make use of the following result (see \cite{Sanguinetti2013} for more details).

    \begin{lemma} \label{LemmaB}
    If there exists a point $z_n$ in $(l_n,u_n)$ such that $f_n^{'} (z_n) = 0 $ with $f_n'(x_n)<0\; \forall x_n \in (l_n, z_n)$ and $f_n'(x_n)>0\; \forall x_n \in (z_n, u_n)$, then it is always $l_n \le x_n^\star  \le z_n$.
\end{lemma}

Using Lemma 2, it follows that solving \eqref{caseA} amounts to looking for the solution of the following equivalent problem:
{\begin{align}\label{caseB}
 \underset{\{x_n\}}{\min} \quad &
                                      \sum\limits_{n=|\mathcal A|+1}^{N} f_n(x_n)\\\nonumber
     \text{subject to} \quad & \sum \limits_{n=|\mathcal A|+1}^{j} x_n\le \rho_j^{\prime} \quad  j=|\mathcal A|+1,\ldots,N\\\nonumber
     & l_n \le x_n \le u^\prime_n \quad  n=|\mathcal A|+1,\ldots,N
\end{align}}
where $u^\prime_n =  z_n$ if $n \in \mathcal{B}$ and $u^\prime_n = u_n$ if $n \notin \mathcal{B}$
with $\mathcal B$ denoting the set of indices {$n$ in \eqref{caseA} } for which case {\bf{b}}) holds true. The above problem is easily seen to be in the same form as \eqref{1} with the only difference that all functions $f_n$ are monotonically decreasing in $(l_n,u'_{n})$ and thus fall into case {\bf{c}}).

The results of Lemmas \ref{LemmaA} and \ref{LemmaB} can be summarized as follows. Once
the optimal values of the variables associated with functions $f_n$ that are monotonically increasing have been trivially computed through Lemma 1, it remains to solve the optimization problem \eqref{caseA} in which the functions $f_n$ belonging to either case {\bf{b}}) or {\bf{c}}). In turn, problem \eqref{caseA} is equivalent to problem \eqref{caseB} with only class {\bf{c}}) functions. This means that we can consider optimization problems of the form in \eqref{1} in which all functions $f_n$ fall into case {\bf{c}}). Accordingly, in the following we assume that \eqref{feasibility} is satisfied and only focus on functions $f_n$ that are continuous, strictly convex and monotonically decreasing, in the intervals $[l_n,u_n]$.

\vspace{-.3cm}
\section{The main result}

This section proposes an iterative algorithm to compute the solutions $x_n^\star$. We begin by calling $h_n(x_n) = -f_n^{'}(x_n)$,
which is a positive and strictly decreasing function since $f_n$ is by definition monotonically decreasing, strictly convex in $[l_n,u_{n}]$ and continuously differentiable in $(l_n,u_{n})$. {We take  $h_n(l_n)=\mathop {\lim }\nolimits_{x_n \rightarrow l_n^+} h_n(x_n)$ and $h_n(u_n)=\mathop {\lim }\nolimits_{x_n \rightarrow u_n^-} h_n(x_n)$.} We also define the functions $\xi_n(\zeta)$ for $n=1,\ldots,N$ as follows
\begin{align}\label{xi_n}
\xi_n(\varsigma)  = \left\{ {\begin{array}{*{20}c}
   {u_n } & {0 \le \varsigma <  h_n (u_n )} \\ \\
   {h_n^{-1}(\varsigma)} & { h_n (u_n ) \le \varsigma  <  h_n(l_n )}\\ \\
      {l_n} & {h_n(l_n ) \le \varsigma}  \\
\end{array}} \right.
\end{align}
where $0 \le \varsigma < +\infty$ and ${h_n^{-1}} $ denotes the inverse function of ${h_n}$ within the interval $[l_n,u_n]$.
Functions $\xi_n(\varsigma) $ can be easily rewritten
in the following compact form:
\begin{align}\label{xi_n.1}
\xi_n(\varsigma) = \min\left\{\max\left\{h_n^{-1}(\varsigma), l_n\right\}, u_n\right\}
\end{align}
from which it is seen that each $\xi_n(\varsigma)$ projects $h_n^{-1}(\varsigma)$ onto the interval $[l_n,u_n]$.

\vspace{-0.3cm}
\begin{theorem}
The solutions of $(\mathcal P)$ are given by
\begin{align}\label{x_n^star}
x_n^\star= \xi_n(\sigma_n^\star) 
\end{align}
where the quantities $\sigma_n^\star$  for  $n=1,\ldots,N$ are some Lagrange multipliers that can be computed by means of the iterative procedure illustrated in {\bf{Algorithm 1}}.\end{theorem}
\vspace{-.3cm}

    \textit{Proof}: Due to space limitations, the proof has been omitted but it can be found in \cite{Sanguinetti2013}. \hfill$\blacksquare$

\begin{algorithm}[t]
\caption{Iterative procedure for solving $(\mathcal P)$ in \eqref{1}.}

\begin{enumerate}\vspace{-0.3cm}
\item Set $j=0$ and $\gamma_n =\rho_n$ for any $n$.
\item {{\bf{While}}} $j < N$
\begin{enumerate}
\item For any $n \in \mathcal{N}_{j}=\{j+1, \ldots, N\}$.
\begin{enumerate}
\item If $\gamma_n < \sum\nolimits_{i=j+1}^n u_i$ then compute $\varsigma_n^\star$ as the solution of
\begin{align}\label{100.10}
c_n(\varsigma)=\sum\nolimits_{i=j+1}^n \xi_i(\varsigma) = \gamma_n.\end{align}

\vspace{-0.3cm}

\item If $\gamma_n \ge \sum\nolimits_{i=j+1}^n u_i$ then set $\varsigma_n^\star=0$.
\end{enumerate}
\item Evaluate
\begin{align}\label{101.10}
\mu^\star &= \underset{n\in \mathcal{N}_{j}}{\max} \quad \varsigma_n^\star \\\label{102.10}
k^\star &= \underset{n\in \mathcal{N}_{j}}{ \max} \quad \left\{n | \varsigma_n^\star=\mu^\star\right\}.
\end{align}

\vspace{-0.4cm}

\item Set $\sigma_n^\star\leftarrow \mu^\star$ for $n=j+1, \ldots,$ $ k^\star$.
\item Use $\sigma_n^\star$ in \eqref{x_n^star} to obtain $x_n^\star$ for $n=j+1, \ldots,$ $ k^\star$.
\item Set $\gamma_n\leftarrow\gamma_n- \gamma_{k^\star}$
for $n=k^\star+1, \ldots, N$.
{\item  Set $j\leftarrow k^\star$.}
\end{enumerate}
\end{enumerate}\vspace{-0.3cm}
\end{algorithm}


As seen, {\bf{Algorithm 1}} proceeds as follows. At the first iteration it sets $j=0$ and $\gamma_n=\rho_n$, $\forall n$, and for those values of $n \in \{1,\ldots,N\}$ such that $\gamma_n < \sum\nolimits_{i=1}^n u_i$
it computes the unique solution $\varsigma_n^\star$ (see \cite{Sanguinetti2013} for more details on existence and uniqueness) of the following equation $c_n(\varsigma) = \gamma_n$
with $c_n(\varsigma) =\sum\nolimits_{i=1}^n \xi_i(\varsigma)$.
On the other hand, for those values of $n\in \{1,2,\ldots,N\}$ such that $\gamma_n \ge \sum\nolimits_{i=1}^n u_i$
it sets $\varsigma_n^\star=0$. The values $\varsigma_n^\star$ computed as described above, for $n=1,\ldots,N$, are first used in \eqref{101.10} and \eqref{102.10} to obtain $\mu^\star$ and $k^\star$, respectively, and then to set $\sigma_n^\star = \mu^\star$ for $n=1,\ldots,k^\star$. Note that if two or more indices can be associated with $\mu^\star$ (meaning that  $\sigma_n^\star = \mu^\star$ for all such indices), then according to \eqref{102.10} the maximum one is selected.
Once $\{\sigma_1^\star,\sigma_2^\star,\ldots,\sigma_{k^{\star}}^\star\}$ have been computed, {\bf{Algorithm 1}} moves to the second step, which essentially consists in solving the reduced problem:\vspace{-0.2cm}
\begin{align}\label{4.1}
     \quad  \underset{\{x_n\}}{\min} \quad &
\sum\limits_{n=k^\star +1}^N f_n(x_n)\\ \nonumber
\text{subject to} \quad & \sum \limits_{n=k^\star +1}^{j} x_n\le \gamma_j - \gamma_{k^{\star}}\quad  j=k^\star +1,\ldots,N\\ \nonumber
 & l_n \le x_n \le u_n \quad  n=k^\star +1,\ldots,N \vspace{-0.1cm}
\end{align}
using the same procedure as before. The procedure terminates in a finite number of steps when all the quantities $\sigma_n^\star$ are computed. The solutions of $(\mathcal P)$ are eventually found as $x_n^\star= \xi_n(\sigma_n^\star)$.
\vspace{-0.3cm}
\subsection{Numerical example and graphical interpretation}
\begin{figure}[t]
\begin{center}
\psfrag{x1}[r][m]{\tiny{$x_1^\star = -0.8$}}
\psfrag{x2}[r][m]{\tiny{$x_2^\star=-1.2$}}
\psfrag{x3}[r][m]{\tiny{$x_3^\star=1.9$}}
\psfrag{x4}[r][m]{\tiny{$x_4^\star=-1.8$}}
\psfrag{x5}[c][m][1][90]{\tiny{$\sigma_3^\star=\sigma_4^\star= 1.195$\quad}}
\psfrag{x6}[c][m][1][90]{\tiny{$\sigma_1^\star=\sigma_2^\star= 4.451$}}
\psfrag{x7}[r][m]{\tiny{$\varsigma$}}
\psfrag{data1}[l][m]{\tiny{$\xi_1(\varsigma)$}}
\psfrag{data2}[l][m]{\tiny{$\xi_2(\varsigma)$}}
\psfrag{data3}[l][m]{\tiny{$\xi_3(\varsigma)$}}
\psfrag{data4}[l][m]{\tiny{$\xi_4(\varsigma)$}}
\includegraphics[width=0.9\columnwidth]{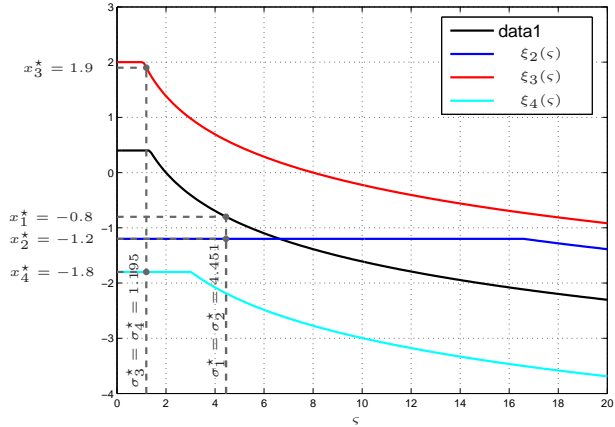}
\end{center}
\caption{Graphical illustration of the solutions $x_n^\star$. The intersection of $\xi_n(\varsigma)$ with the vertical dashed line at $\sigma_n^\star$ yields $x_n^\star$.\vspace{-0.5cm}}
\label{fig1}
\end{figure}

\vspace{-0.1cm}
In the next, we apply the proposed solution to a simple problem and provide graphical interpretations of the general policy spelled out by Theorem 1. For illustration purposes, we assume $N=4$,  $l_n = -\infty$ for any $n$, $\mathbf{u}=[0.4, -1.2, 2, -1.8]$ and ${\boldsymbol{\rho}} = [0.2, -2, 1.1, -1.9]$. In addition, we set $f_n(x_n)= w_n e^{-x_n}$ for any $n$, with $[w_1, w_2, w_3, w_4]=[2, 5, 8, 0.5]$,
from which it follows that $h_n(x_n) ={w_n}e^{-x_n}$ and $h_n^{-1}(\varsigma) = \ln w_n -\ln \varsigma$. Then, from \eqref{xi_n} we obtain\vspace{-0.2cm}
\begin{align}\label{xi_n.0}
\xi_n(\varsigma)  = \left\{ {\begin{array}{*{20}c}
   {u_n } & {0 \le \varsigma  <  w_n e^{-u_n} } \\ \\
   \ln w_n -\ln \varsigma & { w_n e^{-u_n}  \le \varsigma.}\\
\end{array}} \right.
\end{align}
whose graph is shown in Fig. \ref{fig1}.

\begin{figure}[t]
\begin{center}
\psfrag{x1}[r][m]{\tiny{$\gamma_1 = 0.2$}}
\psfrag{x2}[r][m]{\tiny{$\gamma_2 = -2$}}
\psfrag{x3}[r][m]{\tiny{$\gamma_3 = 1.1$}}
\psfrag{x4}[c][m]{\tiny{$ \gamma_4 = -1.9$\quad\quad\quad\quad}}
\psfrag{x6}[c][m][1][90]{\tiny{$\varsigma_1^\star = 1.637$}}
\psfrag{x8}[c][m][1][90]{\tiny{$\varsigma_2^\star = 4.451$}}
\psfrag{x7}[c][m][1][90]{\tiny{$\varsigma_3^\star = 1.196$}}
\psfrag{x5}[c][m][1][90]{\tiny{$\varsigma_4^\star = 2.307$}}
\psfrag{x9}[c][m]{\tiny{$\varsigma$}}
\psfrag{data1}[c][m]{\tiny{$\quad c_1(\varsigma)$}}
\psfrag{data2}[c][m]{\tiny{$\quad c_2(\varsigma)$}}
\psfrag{data3}[c][m]{\tiny{$\quad c_3(\varsigma)$}}
\psfrag{data4}[c][m]{\tiny{$\quad c_4(\varsigma)$}}
\includegraphics[width=0.9\columnwidth]{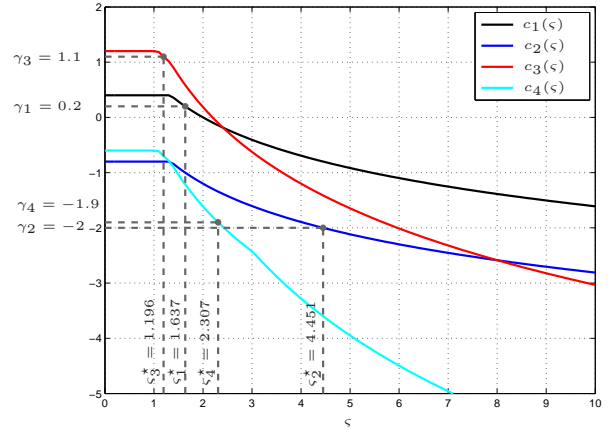}
\end{center}
\caption{Graphical illustration of $c_n(\varsigma)$. Their intersection with the horizontal dashed lines at $\gamma_1 = 0.2$, $\gamma_2 = -2$, $\gamma_3 = 1.1$ and $\gamma_4 = -1.9$ yields respectively $\varsigma_1^\star = 1.637$, $\varsigma_2^\star = 4.451$, $\varsigma_3^\star = 1.196$ and $\varsigma_4^\star = 2.307$. \vspace{-0.4cm}}

\label{fig2}
\end{figure}

As seen, the first operation of {\bf{Algorithm 1}} is to compute the quantities $\varsigma_n^\star$ for $n=1,\ldots, 4$ according to step b1). Since the condition $\gamma_n \le \sum\nolimits_{i=1}^n u_i$ is satisfied for $n=1,\ldots, 4$, the computation of $\varsigma_n^\star$ requires to solve \eqref{100.10} for $n=1,\ldots, 4$. Using \eqref{xi_n.0}, we easily obtain:
\begin{align}
\varsigma_1^\star &= e^{\ln w_1 - \gamma_1} = 1.637\\
\varsigma_2^\star &= e^{{\ln w_1 + u_2 - \gamma_2}} = 4.451\\
\varsigma_3^\star &= e^{{\ln w_3 +  u_1 + u_2 - \gamma_3}} = 1.196\\
\varsigma_4^\star &= e^{\frac{\ln w_1 + \ln w_3  + u_2 + u_4 - \gamma_4}{2}} = 2.307.
\end{align}
%
%
%
A direct depiction of the above results can be easily obtained by plotting $c_n(\varsigma)$ for $n=1,\ldots,4$ as a function of $\varsigma \ge 0$. As shown in Fig. \ref{fig2}, the intersections of curves $c_n(\varsigma)$ with the horizontal lines at $\gamma_n$ yield $\varsigma_n^\star$.

Using the above results into \eqref{101.10} and \eqref{102.10} of step b2) yields $\mu^\star = 4.451$ and $k^\star = 2$
from which (according to step b3)) we obtain $\sigma_1^\star=\sigma_2^\star=\mu^\star = 4.451$. Once the optimal $\sigma_1^\star$ and $\sigma_2^\star$ are computed, {\bf{Algorithm 1}} proceeds solving the following reduced problem:\vspace{-.1cm}
\begin{align}
   \underset{\{x_3,x_4\}}{\min} \quad &
\sum\limits_{n=3}^4 {w_n}e^{-x_n}\\\nonumber
\text{subject to} \quad & \sum \limits_{n=3}^{j} x_n\le \gamma_j\quad  j=3,4\\\nonumber
 & x_n \le u_n \quad  n=3,4
\end{align}
with $\gamma_3 = 3.1$ and $\gamma_4 = 0.1$ as obtained from $\gamma_j \leftarrow \gamma_j - \gamma_{k^\star}$
observing that $\gamma_{k^\star}=\gamma_2 = -2$. Since $\gamma_3  > u_3 $, from step b1) we have that $\varsigma_3^\star = 0$ while $\varsigma_4^\star$ turns out to be given by\vspace{-.1cm}
\begin{align}
\varsigma_4^\star &= e^{{\ln w_3 + u_4 - \gamma_4}} = 1.195.\vspace{-.2cm}
\end{align}
As before, $\varsigma_4^\star$ can be obtained as the intersection of new function $c_4(\varsigma) = \sum\nolimits_{n=3}^4 \xi_n(\varsigma)$ with the horizontal line at $\gamma_4 = 0.1$. Then, from \eqref{101.10} and \eqref{102.10}, we have that $\mu^\star = \mathop {\max }\nolimits_{n=3,4 } \varsigma_n^\star= 1.195$ and $k^\star = 4$. This means that $\sigma_3^\star= \sigma_4^\star=1.195$.

The optimal $x_n^\star$ are eventually obtained as $x_n^\star = \xi_n(\sigma_n^\star)$. This yields $x_1^\star = -0.8$, $x_2^\star = -1.2$, $x_3^\star = 1.9$ and $x_4^\star = -1.8$. As depicted in Fig. \ref{fig1}, the solution $x_n^\star$ correspond to the interception of $\xi_n(\varsigma)$ with the vertical line at $\sigma_n^\star$.
\vspace{-0.3cm}
\subsection{Remarks}


\indent 1) It is worth observing than in deriving {\bf{Algorithm 1}} we have implicitly assumed that the number of linear constraints in \eqref{1} is exactly $N$. When this does not hold true, {\bf{Algorithm 1}} can be slightly modified in an intuitive and straightforward manner. Specifically, let $\mathcal{L} \subset \{1,2,\ldots,N\}$ denote the subset of indices associated to the linear constraints of the optimization problem at hand. The solution of \eqref{1} can still be computed through the iterative procedure illustrated in {\bf{Algorithm 1}} once the two following changes are made: step {\bf{b1}}) replace $\mathcal{N}_{j}$ with $\mathcal{N}_{j} \cap \mathcal{L}$; step {\bf{b5}}) replace the statement ``Set $\gamma_n\leftarrow\gamma_n- \gamma_{k^\star}$
     for $n=k^\star+1, \ldots, N$" with ``Set $\gamma_n\leftarrow\gamma_n- \gamma_{k^\star}$
     for $n \in \{k^\star+1, \ldots, N\} \cap \mathcal{L}$".
As seen, when only a subset $\mathcal{L}$ of constraints must be satisfied, then {\bf{Algorithm 1}} proceeds computing the quantities $\varsigma_n^\star$ only for the indices $n\in \mathcal{L}$.\\
\indent 2) At any given iteration, {\bf{Algorithm 1}} requires solving at most $N-k^\star$ non-linear equations (where $k^\star$ is the value obtained from \eqref{102.10} at the previous iteration):\vspace{-.1cm}
\begin{align}\label{eq1R1}
c_n({\varsigma})= \sum\limits_{i=k^\star+1}^n \xi_i(\varsigma) = \gamma_n
\end{align}\vspace{-.1cm}
for $n=k^\star+1,k^\star+2,\ldots, N$.
When the solutions $\{\varsigma_{n}^\star\}$ of the above equations can be computed in closed form, the computational complexity required by each iteration is nearly negligible. On the other hand, when a closed-form does not exist, this may result in excessive computation. In this latter case, a possible means of reducing the computational complexity relies on the fact that $c_n(\varsigma)$ is a non-increasing function as it is the sum of non-increasing functions. Now, assume that the solution of \eqref{eq1R1} has been computed for $n=n'$. Since we are interested in  the maximum between the solutions of \eqref{eq1R1}, as indicated in \eqref{101.10}, then  for $n''>n'$  $c_{n''}(\varsigma) = \gamma_{n''}$ must be solved only if $c_{n''}(\varsigma_{n'}^\star) > \gamma_{n''}$. Indeed, only in this case $\varsigma_{n''}^\star$ would be greater than $\varsigma_{n'}^\star$. Accordingly, we may proceed as follows. We start by solving \eqref{eq1R1} for $n=k^\star+1$. Then, we look for the first index $n>k^\star+1$ for which $c_{n}(\varsigma_{k^\star+1}^\star) > \gamma_{n}$ and solve the equation associated to such an index. We proceed in this way until $n=N$. In this way, the number of non-linear equations solved at each iteration is smaller than or equal to that required by {\bf{Algorithm 1}}.

\vspace{-0.3cm}
\section{Conclusions}

An iterative algorithm has been proposed to compute the solution of separable convex optimization problems with linear and box constraints. It is particularly interesting since a large number of problems in signal processing and communications can be put in this form, and thus can be efficiently solved with the proposed algorithm.

\vspace{-0.2cm}
\bibliographystyle{IEEEtran}

\end{document}